\font\mybb=msbm10 at 12pt
\def\bb#1{\hbox{\mybb#1}}
\def\Z {\bb{Z}}
\def\R {\bb{R}}
\def\E {\bb{E}}

\documentstyle[12pt]{article}

\catcode`\@=11
\long\def\@makefntext#1{
\protect\noindent \hbox to 3.2pt {\hskip-.9pt
$^{{\ninerm\@thefnmark}}$\hfil}#1\hfill}		

\def\@makefnmark{\hbox to 0pt{$^{\@thefnmark}$\hss}}  
\def\ps@myheadings{\let\@mkboth\@gobbletwo
\def\@oddhead{\hbox{}
\rightmark\hfil\ninerm\thepage}
\def\@oddfoot{}\def\@evenhead{\ninerm\thepage\hfil
\leftmark\hbox{}}\def\@evenfoot{}
\def\sectionmark##1{}\def\subsectionmark##1{}}

\renewcommand{\thefootnote}{\fnsymbol{footnote}}

\newcounter{sectionc}\newcounter{subsectionc}\newcounter{subsubsectionc}
\renewcommand{\section}[1] {\vspace*{0.6cm}\addtocounter{sectionc}{1}
\setcounter{subsectionc}{0}\setcounter{subsubsectionc}{0}\noindent
	{\normalsize\bf\thesectionc. #1}\par\vspace*{0.4cm}}
\renewcommand{\subsection}[1] {\vspace*{0.6cm}\addtocounter{subsectionc}{1}
	\setcounter{subsubsectionc}{0}\noindent
	{\normalsize\it\thesectionc.\thesubsectionc. #1}\par\vspace*{0.4cm}}
\renewcommand{\subsubsection}[1]
{\vspace*{0.6cm}\addtocounter{subsubsectionc}{1}
	\noindent {\normalsize\rm\thesectionc.\thesubsectionc.\thesubsubsectionc.
	#1}\par\vspace*{0.4cm}}

\newcounter{appendixc}
\newcounter{subappendixc}[appendixc]
\newcounter{subsubappendixc}[subappendixc]

\renewcommand{\appendix}[1] {\vspace*{0.6cm}
        \refstepcounter{appendixc}
        \setcounter{figure}{0}
        \setcounter{table}{0}
        \setcounter{equation}{0}
        \renewcommand{\thefigure}{\Alph{appendixc}.\arabic{figure}}
        \renewcommand{\thetable}{\Alph{appendixc}.\arabic{table}}
        \renewcommand{\theappendixc}{\Alph{appendixc}}
        \renewcommand{\theequation}{\Alph{appendixc}.\arabic{equation}}
        \noindent{\bf Appendix \theappendixc #1}\par\vspace*{0.4cm}}

\def\abstracts#1{{
\centering{\begin{minipage}{12.2truecm}\vspace*{.1cm}
        \footnotesize\baselineskip=12pt\noindent
	\parindent=0pt #1
	\end{minipage}}\par}}

\newcounter{itemlistc}
\newcounter{romanlistc}
\newcounter{alphlistc}
\newcounter{arabiclistc}

\newcommand{\fcaption}[1]{
        \refstepcounter{figure}
        \setbox\@tempboxa = \hbox{\footnotesize Fig.~\thefigure. #1}
        \ifdim \wd\@tempboxa > 6in
           {\begin{center}
        \parbox{6in}{\footnotesize\baselineskip=12pt Fig.~\thefigure. #1}
            \end{center}}
        \else
             {\begin{center}
             {\footnotesize Fig.~\thefigure. #1}
              \end{center}}
        \fi}

\newcommand{\tcaption}[1]{
        \refstepcounter{table}
        \setbox\@tempboxa = \hbox{\footnotesize Table~\thetable. #1}
        \ifdim \wd\@tempboxa > 6in
           {\begin{center}
        \parbox{6in}{\footnotesize\baselineskip=12pt Table~\thetable. #1}
            \end{center}}
        \else
             {\begin{center}
             {\footnotesize Table~\thetable. #1}
              \end{center}}
        \fi}

\def\@citex[#1]#2{\if@filesw\immediate\write\@auxout
	{\string\citation{#2}}\fi
\def\@citea{}\@cite{\@for\@citeb:=#2\do
	{\@citea\def\@citea{,}\@ifundefined
	{b@\@citeb}{{\bf ?}\@warning
	{Citation `\@citeb' on page \thepage \space undefined}}
	{\csname b@\@citeb\endcsname}}}{#1}}

\newif\if@cghi
\def\cite{\@cghitrue\@ifnextchar [{\@tempswatrue
	\@citex}{\@tempswafalse\@citex[]}}
\def\citelow{\@cghifalse\@ifnextchar [{\@tempswatrue
	\@citex}{\@tempswafalse\@citex[]}}
\def\@cite#1#2{{$\null^{#1}$\if@tempswa\typeout
	{IJCGA warning: optional citation argument
	ignored: `#2'} \fi}}

 1
 1
 1

\font\ninerm=cmr9

\begin{document}

\centerline{\normalsize\bf $P$-BRANE DEMOCRACY}

\centerline{\footnotesize P.K. TOWNSEND}
\baselineskip=13pt
\centerline{\footnotesize\it DAMTP, University of Cambridge,
Silver St.}
\baselineskip=12pt
\centerline{\footnotesize\it Cambridge, England}
\centerline{\footnotesize E-mail: pkt10@amtp.cam.ac.uk}

\vspace*{0.9cm}
\abstracts{The ten or eleven dimensional origin of central charges in the N=4
or
N=8 supersymmetry algebra in four dimensions is reviewed: while some have a
standard Kaluza-Klein interpretation as momenta in compact dimensions, most
arise from $p$-form charges in the higher-dimensional supersymmetry algebra
that are carried by $p$-brane `solitons'. Although $p=1$ is singled out by
superstring perturbation theory, U-duality of N=8 superstring
compactifications implies a complete `$p$-brane democracy' of the full
non-perturbative theory. An `optimally democratic' perturbation theory is
defined to be one in which the perturbative spectrum includes all particles
with zero magnetic charge. Whereas the heterotic string is optimally
democratic in this sense, the type II superstrings are not, although the
11-dimensional supermembrane might be.}

\normalsize\baselineskip=15pt
\setcounter{footnote}{0}
\renewcommand{\thefootnote}{\alph{footnote}}
\bigskip
\bigskip

Soon after the advent of four-dimensional (D=4) supersymmetry, it was pointed
out (Haag et al. 1975) that the $N$-extended supersymmetry algebra admits a
central extension with $N(N-1)$ central charges: if $Q_\alpha^i$,
$(i=1\dots,N)$, are  the $N$ Majorana-spinor supersymmetry charges and $P_\mu$
the 4-momentum, then
\begin{equation}
\{Q_\alpha^i,Q_\beta^j\} =\delta^{ij}(\gamma^\mu C)_{\alpha\beta} P_\mu +
U^{ij}(C)_{\alpha\beta} + V^{ij}(C\gamma_5)_{\alpha\beta}\ ,
\label{eq:onea}
\end{equation}
where $U^{ij}=-U^{ji}$ and $V^{ij}=-V^{ji}$ are the central charges and $C$ is
the (antisymmetric) charge conjugation matrix. At first, the possibility of
central charges was largely ignored. One reason for this is that the initial
emphasis was naturally on $N=1$ supersymmetry, for which there are no central
charges. Another reason is  that the emphasis was also on massless field
theories; since the central charges appearing in (\ref{eq:onea}) have dimension
of mass they cannot be carried by any massless particle.  Central charges
acquired importance only when massive excitations of extended supersymmetric
theories came under scrutiny. One way that massive excitations naturally arise
is when a gauge group of an otherwise massless gauge theory is spontaneously
broken by vacuum expectation values of scalar fields. Consider $N=4$ super
Yang-Mills (YM) theory with gauge group $SU(2)$ spontaneously broken to
$U(1)$ by a non-vanishing vacuum expectation value of one of the six
(Lie-algebra valued) scalar fields. This can be viewed as an $N=4$
super-Maxwell
theory coupled to a massive $N=4$ complex vector supermultiplet. Clearly, this
massive supermultiplet has maximum spin one, but all massive representations of
the `standard' $N=4$ supersymmetry algebra contain fields of at least spin 2.
The resolution  of this puzzle is that the massive supermultiplet is a
representation of the centrally-extended supersymmetry algebra, which has short
multiplets of maximum spin one. The central charge is the $U(1)$ electric
charge. Similar considerations apply to the magnetic monopole solutions of
spontaneously broken gauge theories with N=4 supersymmetry (Witten and Olive
1978, Osborn 1979). Fermion zero modes in the presence of a magnetic monopole
ensure that the states obtained by semi-classical quantization fall into
supermultiplets.  Since there cannot be bound states of spin greater than one
in any field  theory with maximum spin one (Weinberg and Witten 1980), these
supermultiplets must have maximum spin one and must therefore carry a central
charge, which is in fact the magnetic charge.

The appearance of central charges has usually been seen as a relic of
additional
compact dimensions. Consider $N=2$ supersymmetry, for which there can be two
central charges; to be specific, consider an N=2 super-YM theory with gauge
group $SU(2)$ spontaneously broken to $U(1)$. The electric and magnetic charges
associated with the unbroken $U(1)$ group can be interpreted as momenta in two
extra dimensions, consistent with the natural interpretation of the $N=2$
super-YM theory as a dimensionally-reduced six-dimensional super-YM theory.
However, this cannot be the whole story because this interpretation of central
charges fails when we consider $N>2$. For example, the $N=4$ super-YM theory
can be obtained by dimensional reduction from ten dimensions. If we consider
the momentum in each extra dimension above four as a possible central charge in
the four-dimensional $N=4$ supersymmetry algebra then we find a total of 6
central charges. But the total number of possible central charges is 12, not 6.
One's first reaction to this discrepancy might be to suppose that it is due to
central charges that are already present in the D=10 supersymmetry algebra,
but the full N=1 D=10 super-Poincar\'e algebra does not admit central
charges. It might therefore seem that no D=10 interpretation can be given
to 6 of the 12 central charges in the D=4 N=4 supersymmetry algebra. What this
argument overlooks is that central charges in D=4 might arise from charges that
are {\it not} central in D=10. It is possible (in various spacetime dimensions)
to include $p$-form charges that are central with respect to the
supertranslation algebra but not with respect to the full super-Poincar\'e
algebra (van Holten and Van Proeyen 1982). Together with the momenta in the
extra dimensions, these $p$-form charges provide the higher-dimensional origin
of all central charges in N-extended supersymmetry algebras (Abraham and
Townsend 1991).

For the case under discussion, the 6 `surplus' central charges have their D=10
origin in a self-dual five-form central charge, $Z^+$, in the N=1 D=10
supertranslation algebra:
\begin{equation}
\{Q_\alpha,Q_\beta\} = \big({\cal P}\Gamma^MC\big)_{\alpha\beta}P_M +
\big({\cal P}\Gamma^{MNPQR}C\big)_{\alpha\beta}\, Z^+_{MNPQR}\ ,
\label{eq:oneb}
\end{equation}
where ${\cal P}$ is a chiral projector. Such $p$-form charges are excluded
by the premises of the theorem of Haag, Lopuszanski and Sohnius, and so are not
relevant to representations of supersymmetry by particle states. They are
relevant to extended objects however; in the presence of a $p$-dimensional
extended object, or $p$-brane, the supertranslation algebra aquires a $p$-form
central charge (Azc\'arraga et al. 1989). We might therefore expect to find
a fivebrane solution of D=10 super-YM theory corresponding to the five-form
charge in (\ref{eq:oneb}) and there is indeed such a solution. It is found by
interpreting the YM instanton of four dimensional Euclidean space as an
`extended soliton' in ten dimensional Minkowski space (Townsend 1988).
By the inverse construction, which may be interpreted as `wrapping' the
fivebrane soliton around a 5-torus, one obtains an `instantonic' soliton in
five
dimensions, i.e. a solution of the self-duality condition for YM fields on
$R^4$. By solving these equations on $R^3\times S^1$, which amounts to a
further
$S^1$ compactification to four spacetime dimensions, we  obtain a BPS monopole
of the four-dimensional YM/Higgs equations. The magnetic charge it carries is
one of six possible types because there are six Higgs fields in the $N=4$
super-YM multiplet. Which one gets an expectation value depends on which
5-torus is chosen for the first step in the construction, and this can be done
in six ways. Thus, six central charges are momenta in the six extra
dimensions but six more arise from the five-form charge in ten dimensions. From
this interpretation, it is clear that we should expect to get `surplus' central
charges only for spacetime dimension $D\le5$ because only in these cases can a
fiveform charge yield a scalar charge on dimensional reduction. As a check,
consider the N=4 super YM theory in, say, D=7; the relevant
supersymmetry algebra has an $SO(3)$ automorphism group, and the central
charges
in the relevant supersymmetry algebra belong to a ${\bf 3}$ of $SO(3)$, i.e.
there are three of them, just the number of extra dimensions. In D=5, the
relevant supersymmetry algebra has a $USp(4)$ automorphism group and central
charges belong to a ${\bf 5}\oplus {\bf 1}$ of $USp(4)$, i.e. there are five
central charges for the five extra dimensions and one from dimensional
reduction of the fiveform charge in ten dimensions.  As a final point, observe
that since the LHS of (\ref{eq:oneb}) is a $16\times 16$ real symmetric matrix
the maximal number of algebraically distinct charges that can appear on the RHS
is 136, which is precisely the total number of components of
$P_M$ and $Z^+_{MNPQR}\, $.

Many of the observations made above concerning super YM theories can be
generalized to supergravity theories. Certain N=2 and N=4 supergravity theories
can be considered as compactifications of D=6 and D=10 supergravity on $T^2$
and $T^6$, respectively. The D=10 case is of particular interest because of its
close connection with the heterotic string. Consider first D=10 supergravity
coupled to a rank $r$ semi-simple D=10 super-YM theory. The generic four
dimensional massless field theory resulting from a compactification on $T^6$ is
N=4 supergravity coupled to (6+$r$) abelian N=4 vector supermultiplets, six of
which contain the Kaluza-Klein (KK) gauge fields for $U(1)^6$, the isometry
group of $T^6$. Since the graviton multiplet contains six vector fields, the
total gauge symmetry group is $U(1)^{12+r}$ and the corresponding field
strengths can be assigned to the irreducible vector representation of
$SO(6,6+r)$, which is a symmetry group of the four-dimensional effective
action. The massive excitations in four dimensions discussed above for the pure
super-YM theory are still present, since in addition to the massive vector
mutiplets arising from the breaking of the rank $r$ gauge group to $U(1)^r$,
there are also gravitational analogues of the BPS monopoles (Harvey and Liu
1991, Gibbons et al. 1994, Gibbons and Townsend 1995), which can be viewed
(Khuri 1992, Gauntlett et al. 1993) as `compactifications' of Strominger's
fivebrane solution (Strominger 1990) of the D=10 heterotic string. These
excitations are sources for the $U(1)^r$ fields. However, there are now new
massive excitations of KK origin; the electrically charged excitations arise
from the harmonic expansion of the fields on $T^6$, while the magnetically
charged ones are the states obtained by semi-classical quantization of KK
monopoles. These serve as sources for the six KK gauge fields. This provides us
with massive excitations carrying $2\times (6+r)$ of the $2\times (12+r)$ types
of electric and magnetic charges.  What about massive excitations carrying the
remaining 6+6 electric and magnetic charges? There are none in the context of
pure KK theory but the electrically charged states, at least, are present in
string theory; they are the string winding modes around the six-torus. In D=10
the magnetic dual to a string is a fivebrane, so we should expect to find the
corresponding magnetically charged states as compactifications on $T^6$ of a
new D=10 fivebrane of essentially gravitational origin. Such a solution indeed
exists. In terms of the string metric, the bosonic action of D=10 N=1
supergravity is
\begin{equation}
S= \int\! d^{10}x \, e^{-2\phi}\big[ R + 4(\nabla \phi)^2 - {1\over3}H^2\big]
\label{eq:onec}
\end{equation}
where $H$ is the three-form field strength of the two-form potential that
couples to the string, and $\phi$ is the scalar `dilaton' field. The field
equations have a `neutral' fivebrane solution (Duff and Lu 1991, Callan et al.
1991) for which the metric is
\begin{equation}
ds^2 = -dt^2 + d{\bf y}\cdot d{\bf y} + \Bigg[1+ {\mu_5 \over \rho^2}\Bigg]
\Big(d\rho^2 + \rho^2d\Omega_3^2\Big) \ ,
\label{eq:oned}
\end{equation}
where ${\bf y}$ are coordinates for $\E^5$, i.e. the fivebrane is aligned
with the ${\bf y}$-axes, $d\Omega_3^2$ is the $SO(4)$-invariant metric on the
three-sphere, and $\mu_5$ is an arbitrary constant (at least in the classical
theory; we shall return to this point below). This solution is geodesically
complete because $\rho=0$ is a null hypersurface at infinite affine parameter
along any geodesic. Clearly, this solution of D=10 supergravity is also a
solution of D=10 supergravity coupled to a D=10 super-YM theory because the YM
fields vanish (hence the terminology `neutral'). It may therefore be considered
as an approximate solution of the heterotic string theory, but it is not an
exact solution because the effective field theory of the heterotic string
includes additional interaction terms involving the Lorentz Chern-Simons
three-form. There is an exact solution, the `symmetric' fivebrane, that takes
account of these terms (Callan et al. 1991). Since the metric of the symmetric
fivebrane solution is identical to that of the neutral fivebrane, it seems
possible that, in the string theory context, the latter should be simply
replaced by the former. In any case, the difference will not be of importance
in this contribution. The main point for the present discussion is that there
is an additional fivebrane in the supergravity context whose `wrapping modes'
are magnetically charged particles in D=4 that are the magnetic duals of the
string winding modes.

We have now found massive excitations carrying all 28+28 electric and magnetic
charges of the heterotic string (since $r=22$ in this case). Because of the
Dirac quantization condition, only 28 of the 56 possible types of charges can
be
carried by states in the perturbative spectrum, no matter how we choose the
small parameter of perturbation theory. Heterotic string theory is `optimal' in
the sense that states carrying 28 different types of electric charge appear in
string perturbation theory, whereas perturbative KK theory has states carrying
only 22 electric charges. However, one can try to `improve' KK theory by taking
into account the fact that D=10 supergravity admits the extreme string solution
(Dabholkhar et al. 1990) for which the metric is
\begin{equation}
ds^2 = \Bigg[1+ {\mu_1\over \rho^6}\Bigg]^{-1}  \Big(-dt^2 + d\sigma^2\Big) +
\Big(d\rho^2 + \rho^2d\Omega_7^2\Big) \ ,
\label{eq:onee}
\end{equation}
where $\sigma$ is one of the space coordinates, i.e. the string is aligned with
the $\sigma$-axis, $d\Omega_7^2$ is the $SO(8)$-invariant metric on the
seven-sphere, and $\mu_1$ is an arbitrary constant proportional to the string
tension. Winding modes of this string `soliton' carry the extra 6 electric
charges that are missing from the KK theory. This suggests that we bring the
extra 6 charges into perturbation theory by identifying this solitonic string
with a fundamental string.

There is a suggestive analogy here with the Skyme model of baryons: in the
limit of vanishing quark masses the pions are the massless fields of QCD and
the non-linear sigma model their effective action, just as D=10 supergravity
can be seen as the effective action for the massless modes of a string theory.
The sigma-model action has Skyrmion solutions that carry a topological charge
which can be identified as baryon number, and these solutions  are {\it
identified} with the baryons of QCD. A potential difficulty in the D=10
supergravity case is that the string solution (\ref{eq:onee}) is not really
solitonic because the singularity at $\rho=0$ is a naked timelike
singularity at finite affine parameter. This could of course be taken simply as
a further indication that one should introduce the string as a fundamental one
and relinquish any attempt to interpret the solution (\ref{eq:onee}) as a
soliton. On the other hand, there is a similar difficulty in the pion sigma
model: there are no non-singular static solutions carrying baryon number unless
an additional, higher-derivative, term is included in the action. Perhaps
something similar occurs in D=10 supergravity.

Another potential difficulty in trying to interpret the string soliton
(\ref{eq:onee}) as  the fundamental string is that the constant $\mu_1$ in
this solution is arbitrary. The resolution of this difficulty
requires consideration of the quantum theory. First, we observe that the reason
that (\ref{eq:onee}) is called `extreme' is that it saturates a Bogomolnyi-type
lower bound on the string tension in terms of the charge $q_e= \oint\!
e^{-2\phi}\star H$, where $\star$ is the Hodge dual of $H$ and the integral is
over the  seven-sphere at `transverse spatial infinity'. A similar result holds
for the constant $\mu_5$ of the fivebrane solution (\ref{eq:oned}) but with the
`electric' charge $q_e$ replaced by the `magnetic' charge  $q_m= \oint\! H$,
where now the integral is over the three-sphere at transverse spatial infinity.
Given these facts, the parameters $\mu_1$ and $\mu_5$ are proportional, with
definite constants of proportionality whose precise values are not important
here, to $q_e$ and $q_m$ respectively. Second, the existence of a non-singular
magnetic dual fivebrane solution implies a quantization of $q_e$. Specifically,
the product $q_e q_m$ is quantized as a consequence of a generalization of the
Dirac quantization condition in dimension D to $p$-branes, and their duals of
dimension $\tilde p = D-p-4$ (Nepomechie 1985, Teitelboim 1986). Thus, not only
is $q_e$ quantized, but so too is $q_m$.

One way to establish the above mentioned Bogomolnyi bound on the constants
$\mu_1,\mu_5$ is to make use of the supersymmetry algebra. In the fivebrane
case one can relate the charge $q_m$ to a contraction of the five-form charge
in the D=10 algebra (\ref{eq:oneb}) with the five-vector formed from the
outer-product of the five spatial translation Killing vectors of the fivebrane
solution. The Bogomolnyi bound can then be deduced from the supersymmetry
algebra by a procedure modeled on the derivation of Witten and Olive for
magnetic monopoles in D=4. There might appear to be an asymmetry between
strings and fivebranes in this respect because while (\ref{eq:oneb}) includes a
five-form charge, carried by fivebranes, it does not include the corresponding
one-form charge that we might expect to be carried by strings. To include such
a
charge we must modify the algebra (\ref{eq:oneb}) to
\begin{equation}
\{Q_\alpha,Q_\beta\} = \big({\cal P}\Gamma^MC\big)_{\alpha\beta}(P_M + T_M) +
\big({\cal P}\Gamma^{MNPQR}C\big)_{\alpha\beta}Z^+_{MNPQR}\ .
\label{eq:oneh}
\end{equation}
This algebra is isomorphic to the previous one, which is why the absence of the
one-form charge $T$ did not show up in the earlier counting exercise;
classically, $T$ can be absorbed into the definition of $P$. But suppose $k$ is
an  everywhere non-singular spacelike Killing vector of spacetime, with closed
orbits of length $R$; then the eigenvalues of the scalar operators $k\cdot P$
and $k\cdot T$ are multiples of $R^{-1}$ and $R$, respectively, so in the
quantum theory $T$ cannot be absorbed into the definition of $P$. Moreover, it
follows from the form of the Green-Schwarz action for the heterotic superstring
that this one-form charge is indeed present in the algbra (Azc\'arraga et al.
1989, Townsend 1993).

The massive states of the heterotic string carry a total of 28 electric and 28
magnetic charges, as mentioned above, but only 12 linear combinations can
appear as central charges in the N=4 D=4 supersymmetry algebra. This fact leads
to some interesting consequences, e.g. symmetry enhancement at special vacua
(Hull and Townsend 1995b). Here, however, I shall concentrate on theories which
have N=8 supergravity as their effective D=4 field theory, e.g. type II
superstrings compactified on a six-torus. In this case, the 56 electric or
magnetic charges associated with the 28 abelian gauge fields of N=8
supergravity {\it all} appear as central charges in the N=8 supersymmetry
algebra. From the standpoint of KK theory, only those massive states carrying
the six electric KK charges appear in perturbation theory. Type II superstring
theory improves on this by incorporating into perturbation theory the string
winding modes, which carry six more electric charges, but this still leaves 16
electric charges unaccounted for. These 16 charges are those which would, if
present in the spectrum, couple to the Ramond-Ramond (R-R) gauge fields of the
D=4 type II string. They are absent in perturbation theory, however, because
the R-R gauge fields couple to the string through their field strengths only.
This has long been recognized as a problematic feature of type II superstrings,
e.g. in the determination of the free type II string propagator (Mezincescu et
al. 1989).

An alternative way to see that states carrying RR charges must be
absent in perturbation theory is to note that the set of 56 electric plus
magnetic central charges can be assigned to the irreducible ${\bf 56}$
representation of the duality group $E_{7,7}$ (Cremmer and Julia 1978,1979),
which becomes the U-duality group $E_7(\Z)$ of the string theory (Hull and
Townsend 1995a). Now $E_{7,7}\supset Sl(2;\R)\times SO(6,6)$, so the U-Duality
group has as a subgroup the product of the S-Duality group $Sl(2;\Z)$ and the
type II T-Duality group $SO(6,6;\Z)$. With respect to $Sl(2;\R)\times SO(6,6)$
the
${\bf 56}$ of $E_{7,7}$ decomposes as
\begin{equation}
{\bf 56} \rightarrow ({\bf 2},{\bf 12}) \oplus ({\bf 1},{\bf 32})\ .
\label{eq:onei}
\end{equation}
The analogous decomposition of the $({\bf 2}, {\bf 28})$ representation of the
$S\times T$ duality group $Sl(2;\R)\times SO(6,22)$ of the generic
$T^6$-compactified heterotic string is
\begin{equation}
({\bf 2}, {\bf 28})\rightarrow ({\bf 2},{\bf 12}) \oplus 16\times ({\bf
2},{\bf 1})\ ,
\label{eq:extra}
\end{equation}
which makes it clear that the $({\bf 1},{\bf 32})$ representation in
(\ref{eq:onei}) is that of the 16+16 electric and magnetic RR charges, which
are
therefore S-Duality inert and transform {\it irreducibly} under T-Duality.
Since
T-Duality is perturbative and magnetic monopoles cannot appear in perturbation
theory, this means that both  electric and magnetic RR charges must be
non-perturbative. Moreover, while the complete absence from the spectrum of
states carrying RR charges would be consistent with S and T duality, their
presence is required by  U-duality. In fact, these states are present (Hull and
Townsend 1995a). They are $p$-brane `wrapping modes' for
$p>1$, which explains their absence in perturbative string theory.

Thus, in contrast to the heterotic string, the type
II string is {\it non-optimal}, in the sense that it does not incorporate into
perturbation theory {\it all} electrically charged states. Just as string
theory
improves on KK theory in this respect, one wonders whether there is some theory
beyond string theory that is optimal in the above sense. As a first step in
this direction one can try to `improve' string theory, as we tried to `improve'
KK theory, by incorporating p-brane solitons that preserve half the D=10 N=2
supersymmetry (no attempt will be made here to consider solitons that break
more than half the supersymmetry). In
order to preserve half the supersymmetry, a $p$-brane must carry a p-form
central charge in the D=10 N=2 supertranslation algebra, so we shall begin by
investigating the possibilities for such p-form charges. Consider first the
N=2A D=10 supersymmetry algebra. Allowing for all algebraically inequivalent
p-form charges permitted by symmetry, we have
\begin{eqnarray}
\label{eq:onej}
\{Q_\alpha,Q_\beta\} &=& \big(\Gamma^MC\big)_{\alpha\beta} P_M
+ (\Gamma_{11}C)_{\alpha\beta}Z + \big(\Gamma^M\Gamma_{11}C\big)_{\alpha\beta}
Z_M + \big(\Gamma^{MN}C\big)_{\alpha\beta}Z_{MN} \nonumber \\
&&+\, \big(\Gamma^{MNPQ}\Gamma_{11}C\big)_{\alpha\beta}Z_{MNPQ}
+ \big(\Gamma^{MNPQR}C\big)_{\alpha\beta}Z_{MNPQR}\ .
\end{eqnarray}
The supersymmetry charges are 32 component non-chiral D=10 spinors, so the
maximum number of components of charges on the RHS is 528, and this maximum is
realized by the above algebra since
\begin{equation}
10 + 1 + 10 + 45 + 210 + 252 = 528\ .
\label{eq:onek}
\end{equation}
Note that in this case the $\Gamma_{11}$ matrix distinguishes between the
term involving the 10-momentum $P$ and that involving the one-form charge
carried by the type IIA superstring. This means that on compactification to
D=4 we obtain an additional six electric central charges from this source,
relative to the heterotic case. These are balanced by an additional six
magnetic charges due to the fact that the five-form charge is no longer
self-dual as it was in the heterotic case. Thus, there is now a total of 24
D=4 central charges carried by particles of KK, string, or fivebrane origin.
These are the charged particles in the NS-NS sector of the superstring theory.

The remaining 32 D=4 central charges of the D=4 N=8 supersymmetry algebra have
their D=10 origin in the zero-form, two-form and four-form charges of the D=10
algebra (\ref{eq:onej}). One might suppose from this fact that these 32 charges
would be carried by particles in the (non-perturbative) R-R sector whose D=10
origin is either a D=10 black hole, membrane or fourbrane solution of the IIA
supergravity theory, but this is only partly correct. There are indeed R-R
$p$-brane solutions of D=10 IIA supergravity for $p=0,2,4$, in addition to the
NS-NS $p$-brane solutions for $p=1,5$, but there is also a R-R $p$-brane
solution for $p=6$. (Horowitz and Strominger 1991). The IIA p-branes with
$p=(0,6)\,;\, (1,5)\, ;\, (2,4)$ are the (electric, magnetic) sources for the
one-form, two-form and three-form gauge fields, respectively, of type IIA
supergravity.

The reason for this mis-match can be traced to the fact that the
four-form charge in (\ref{eq:onej}) actually contributes `twice' to the D=4
central charges because apart from the obvious 15 charges it also contributes
an additional one central charge via the component $Z_{\mu\nu\rho\sigma}$,
which is equivalent to a scalar in D=4. This scalar can alternatively be viewed
as the obvious scalar charge in D=4 associated with a six-form charge in the
D=10 algebra. However, a six-form charge is algebraically equivalent to a
four-form charge, which explains why it is absent in (\ref{eq:onej}) and why it
is not needed to explain the 56 D=4 central charges. It is possible that the
situation here for the six-form charge in the IIA algebra is analogous to the
one-form charge in the heterotic case in that it may be necessary to include it
as a separate charge in the quantum theory even though it is not algebraically
independent of the other charges.

One objection that can be made to the association of $p$-brane solutions of
a supergravity theory with $p$-form charges in the supersymmetry
algebra is that the possibility just noted of replacing a four form by a six
form in D=10 is of general applicability. Thus, a $p$-form charge in a
$D$-dimensional supersymmetry algebra could always be replaced by an
algebraically equivalent $(D-p)$-form charge. Note that this is not simply a
matter of exchanging one $p$-brane for its dual because the dual object is a
$(D-p-4)$-brane, not a $(D-p)$-brane. As a result of this ambiguity, one cannot
deduce from the algebra alone which $p$-brane solutions will occur as
solutions of the supergravity theory; one needs additional information.
Fortunately, this information is always available, and it is always the case
that the supersymmetry algebra admits a $p$-form charge whenever the
supergravity multiplet contains a $(p+1)$-form potential.

Let us now consider how this type of analysis fares when applied to the type
IIB
superstring. The IIB supersymmetry algebra has two Majorana-Weyl supercharges,
$Q_\alpha^i$, (i=1,2), of the same chirality and,  allowing for p-form central
charges, the supertranslation algebra is
\begin{eqnarray}
\label{eq:onel}
\{Q_\alpha^i,Q_\beta^j\} &=\, \delta^{ij}\big({\cal
P}\Gamma^MC\big)_{\alpha\beta} P_M + \big({\cal
P}\Gamma^MC\big)_{\alpha\beta}\,
\tilde Z_M^{ij} + \varepsilon^{ij}\big({\cal
P}\Gamma^{MNP}C\big)_{\alpha\beta}\, Z_{MNP}\nonumber \\
& +\delta^{ij} \big({\cal P}\Gamma^{MNPQR}C\big)_{\alpha\beta}(Z^+)_{MNPQR}+
\big({\cal P}\Gamma^{MNPQR}C\big)_{\alpha\beta}(\tilde Z^+)_{MNPQR}^{ij}\; ,
\end{eqnarray}
where the tilde indicates the tracefree symmetric tensor of
SO(2), equivalently a $U(1)$ doublet. The total number of components of all
charges on the RHS of (\ref{eq:onel}) is
\begin{equation}
10 + 2\times 10 + 120 + 126 + 2\times 126 = 528\ .
\label{eq:onem}
\end{equation}
Moreover, {\it all} $p$-form charges are needed to provide a D=10
interpretation of the 56 central charges of the D=4 N=8 supersymmetry algebra.
I emphasize this point because it is not what one might expect given that
D=10 IIB supergravity admits $p$-brane solutions for $p=1,3,5$, with there
being two strings and two fivebranes because there are two two-form gauge
potentials. Each of these $p$-brane solutions can be paired with a $p$-form
charge in the supersymmetry algebra, but this leaves one self-dual
five-form charge without an associated fivebrane solution. This is because
there are {\it three} (self-dual) five-form charges, not two.

Perhaps even more surprising is that origin of this discrepancy lies in the
NS-NS sector and not in the R-R sector. The reason for this has to do with the
D=10 interpretation of the magnetic duals to the 6 KK charges, i.e. the
magnetic charges carried by the KK monopoles. In type II string theory the KK
charges are related by T-duality to the string winding modes. These have their
origin in the D=10 NS-NS string for which the magnetic dual is a fivebrane.
This
fivebrane is associated with a five-form charge, so T-duality implies that the
duals of the KK charges also have their D=10 origin in a fiveform charge. This
is true for both the type IIA and the type IIB superstrings. In the type IIA
case there was apparently only one five-form charge in the supesymmetry algebra
but, in distinction to the heterotic case, it was not self dual. Thus,
effectively there were two five-form charges: the self-dual one, in common with
the heterotic string, and an additional anti-self-dual one. The additional one
is {\it not} associated with a $p$-brane solution of the D=10 supergravity
theory but, instead, is associated with the KK monopoles. From this perspective
it is not surprising that there are three, rather than two, self-dual five-form
charges in the type IIB supersymmetry algebra.

Having established the potential importance of $p$-brane solutions of the D=10
supergravity theories, the next step is to determine whether these solutions
are
non-singular. The NS-NS string is singular but it may be identified with the
fundamental string and, as noted earlier, the NS-NS fivebrane solution
(\ref{eq:oned}) is geodesically complete, so we need concern ourselves only
with the additional R-R $p$-branes. In the type IIB case these comprise a
string, a threebrane and a  fivebrane. While the threebrane is non-singular
(Gibbons et al. 1995), the R-R string and fivebrane are singular (Townsend
1995a, Hull 1995), and the significance of this is unclear at present. In the
type IIA case, the R-R $p$-branes comprise a zero-brane, i.e. extreme `black
hole', a two-brane, i.e. membrane, a four-brane and a six-brane. Again, all are
singular but in this case there is a simple resolution of this difficulty,
which we shall explain shortly.

At this point it should be clear that, for either the type IIA or type IIB
superstring compactified on $T^6$, one can account for the existence of states
in the spectrum carrying all 56 charges provided account is taken of the
wrapping modes of the D=10 p-brane solitons associated with the p-form charges
in the D=10 supersymmetry algebra. From the standpoint of perturbative string
theory, $p=1$ is a special value since string theory incorporates $p=1$
wrapping
modes, alias string winding modes, into perturbation theory. However, U-Duality
of the non-perturbative D=4 string theory implies that the distinction between
$p=1$ and $p>1$ is meaningless in the context of the full non-perturbative
theory: i.e. {\it U-Duality implies a complete $p$-brane `democracy'}, hence
the title of this contribution. It is merely by convention that we continue to
refer to this non-perturbative theory as `string' theory.

Having just said, in effect, that `all p-branes are equal', perhaps we can
nevertheless allow ourselves the luxury of considering, following Orwell's
dictum, that some are more equal than others. Specifically, it is convenient to
divide the 56 central charges into the electric ones and the magnetic ones. I
have emphasized above that the heterotic string is `optimal' in that it
incorporates all electric charges into perturbation theory; we might now say
that it is `as democratic' as a perturbative theory can be. Is there a
similarly `optimally democratic' perturbative theory underlying the type II
string theories? The answer is a qualified `yes', at least in the type IIA
case, and it involves consideration of D=11 supergravity, to which we now turn
our attention.

D=11 supergravity compactified on $T^7$ has in common with the type II
superstrings compactified on $T^6$ that the effective D=4 field theory is N=8
supergravity (Cremmer and Julia 1978,1979). From the D=11 standpoint, seven of
the 56 central charges can be interpreted as momenta in the extra dimensions,
i.e. as KK electric charges, but this still leaves 49 unaccounted for. These
remaining 49 charges must have a D=11 origin as p-form charges. Allowing for
all possible p-form charges, the D=11 supersymmetry algebra is
\begin{equation}
\{Q_\alpha,Q_\beta\} = \big(\Gamma^MC\big)_{\alpha\beta}P_M +
\big(\Gamma^{MN}C\big)_{\alpha\beta}\, Z_{MN} +
\big(\Gamma^{MNPQR}C\big)_{\alpha\beta}\, Z_{MNPQR}\ .
\label{eq:onen}
\end{equation}
That is, there is a two-form and a five-form charge. The total number of
components of all charges on the RHS of (\ref{eq:onen}) is
\begin{equation}
11 + 55 + 462 = 528\ ,
\label{eq:oneo}
\end{equation}
which is, algebraically, the maximum possible number. From this algebra one
might guess that D=11 supergravity admits p-brane solutions that preserve half
the supersymmetry for p=2 and p=5, and this guess is correct (Duff and Stelle
1991, G\"uven 1992). In the KK theory the only charged massive states are the
KK modes carrying 7 of the 28 electric charges and the KK monopoles carrying
the corresponding 7 magnetic charges. The remaining 21 electric charges are
carried by the wrapping modes of the D=11 twobrane, i.e. membrane, while the
corresponding 21 magnetic charges are carried by wrapping modes of the D=11
fivebrane. Note that all magnetic charges have their D=11 origin in the
five-form charge; as for the type IIA string the five-form charge accounts not
only for the fivebrane charges but also for the charges carried by KK
monopoles.

The 11-metric for both the membrane and the fivebrane solution of
D=11 supergravity can be written as
\begin{equation}
ds^2 = \Bigg[1+ {\mu_p\over \rho^{(8-p)}}\Bigg]^{-{2\over (p+1)}}(-dt^2 + d{\bf
y}\cdot d{\bf y}) + \Bigg[1+ {\mu_p\over
\rho^{(8-p)}}\Bigg]^{2\over (8-p)}\Big(d\rho^2 +
\rho^2d\Omega_{(9-p)}^2\Big) \ ,
\label{eq:onep}
\end{equation}
where ${\bf y}$ are coordinates of $\E^p$, so the p-brane is aligned
with the ${\bf y}$ axes, and $\mu_p$ is a constant. In
both cases there is a singularity at $\rho=0$ and this was originally
interpreted as due to a physical source. However, this singularity is merely a
coordinate singularity, and the hypersurface
$\rho=0$ is an event horizon. Since the horizon can be reached and crossed in
finite proper time, one might think that the appropriate generalization of the
singularity theorems of General Relativity would imply the existence of a
singularity behind the horizon. This is indeed the case for p=2, and the
Carter-Penrose diagram in this case is rather similar to that of the extreme
Reissner-Nordstrom (RN) solution of General Relativity, i.e. a timelike
curvature singularity hidden behind an event horizon (Duff et al. 1994). For
p=5, however, the the analytic continuation of the exterior metric through the
horizon leads to an interior metric that is isometric to the exterior one. The
maximally analytic extension of this exterior metric is therefore geodesically
complete for p=5 (Gibbons et al. 1995). Thus, the fivebrane is genuinely
solitonic while the membrane has a status similar to that of the extreme RN
black hole in GR. This disparity suggests that we identify the membrane
solution as the fields exterior to a {\it fundamental} supermembrane. Various
reasons in favour of this idea can be found in the literature (Hull and
Townsend 1995a, Townsend 1995a 1995b). Another one is that it could allow us to
bring into perturbation theory the 21 electric charges that cannot be
interpreted as momenta in the extra 7 dimensions. Thus, a fundamental
supermembrane theory is `optimal', in the sense that {\it all} electric charges
appear in perturbation theory. This presupposes, of course, that some sense can
be made of supermembrane perturbation theory. Until now, attempts in this
direction have been based on the the worldvolume action for a D=11
supermembrane (Bergshoeff et al. 1987,1988), but this approach runs into the
difficulty that the spectrum is most likely continuous (de Wit et al 1989),
which would preclude an interpretation in terms of particles. We
shall return to this point at the conclusion of this contribution.

This is a convenient point to summarize the p-brane solutions of the N=2 D=10
supergravity theories and of D=11 supergravity via the `type II
Branescan' of Table 1. Note that each $p$-brane has a dual of dimension $\tilde
p = D-p-4$, except the type IIB D=10 threebrane which is self-dual (Horowitz
and
Strominger 1991, Duff and Lu 1991b).

\begin{table}[h]
\tcaption{{\bf The TYPE II Branscan}}
\small
\centering
\begin{tabular}{||c||c|c|c|c|c|c|c||}
\hline\hline
{}{} &{}{} &{}{} &{}{} & {}{} & {}{} & {}{} & {}{} \\
11{} &{}{} &{}{} & 2{} & {}{} & {}{} & 5{} & {}{} \\
{}{} &{}{} &{}{} &{}{} & {}{} & {}{} & {}{} & {}{} \\
\hline
{}{} &{}{} &{}{} &{}{} & {}{} & {}{} & {}{} & {}{} \\
10A&\ \ 0\ \  & 1 {} & \ \ 2\ \ & {}{} & \ \ 4\ \  & 5 {} & \ \ 6 \ \ \\
{}{} &{}{} &{}{} &{}{} & {}{} & {}{} & {}{} & {}{} \\
\hline
{}{} &{}{} &{}{} &{}{} & {}{} & {}{} & {}{} & {}{} \\
10B&{}{} &1+1&{} {} & \ \ 3\ \ & {}{} & 5+5 & {}{} \\
{}{} &{}{} &{}{} &{}{} & {}{} & {}{} & {}{} & {}{} \\
\hline\hline
\end{tabular}
\end{table}

We are now in a position to return, as promised, to the resolution of the
problem in type IIA superstring theory that the RR $p$-brane solutions are
singular. Note first that the IIA supergravity can be obtained by dimensional
reduction from D=11 supergravity, which was how it was first constructed
(Gianni and Pernici 1984, Campbell and West 1984). The Green-Schwarz action
for the type IIA superstring (Green and Schwarz 1984) can be obtained
(Duff et al. 1987) by double dimensional reduction of the worldvolume action
of the D=11 supermembrane. The D=10 extreme string solution is similarly
related to the D=11 membrane solution (Duff and Stelle 1991) and this relation
allows the singularity of the string solution to be reinterpreted as a mere
coordinate singularity at the horizon in D=11 (Duff et al. 1994).
As mentioned above, there remains a further singularity behind the horizon.
It is not clear what the interpretation of this singularity should be. One
can argue that `clothed' singularities are not inconsistent with the soliton
interpretation, as has been argued in the past for extreme RN black holes
(Gibbons 1985), or one can argue that the D=11 membrane solution should be
interpreted as a fundamental supermembrane. It should be noted here that there
are strong arguments against simply discarding the membrane solution: it is
needed for U-duality of the D=4 type II superstring (Hull and Townsend 1995a)
and for the symmetry enhancement in $K_3$-compactified D=11 supergravity
(Hull and Townsend 1995b) needed for the proposed equivalence (Witten, 1995) to
the $T^3$ compactified heterotic string.

Let us now turn to the other $p$-brane solutions of the type IIA theory.
First, the fourbrane can be interpreted as a double dimensional
reduction of the D=11 fivebrane, in which the singularity of the fourbrane
becomes a coordinate singularity at the horizon of the fivebrane (Duff et al.
1994). Thus both the D=10A string and fourbrane in Table 1 are derived by
double
dimensional reduction from the D=11 membrane and fivebrane diagonally above.
Second, the D=10 membrane and D=10 fivebrane can each be interpreted as a
superposition  of the corresponding D=11 solutions, so each of these D=10A
solutions in Table 1 has a straightforward interpretation as the dimensional
reduction of the D=11 solution directly above it. There is, however, an
important distinction between the D=10A membrane and the D=10A fivebrane: in
the membrane case it is {\it necessary} to pass to D=11 to remove the
singularity, whereas this is optional for the fivebrane (as expected from the
fact that this fivebrane solution must do triple purpose as both the type IIA
fivebrane and the heterotic and NS-NS type IIB fivebrane). Third, the
type IIA sixbrane solution can be interpreted as a direct analogue in D=10 of
the Kaluza-Klein monopole in D=4. Just as the latter becomes non-singular in
D=5, so the singular D=10 sixbrane becomes non-singular in D=11 (Townsend
1995a). At this point we may pause to note that all {\it magnetic} $p$-brane
solutions have now been interpreted as {\it completely non-singular} solutions
in D=11.

This leaves the electric D=10A 0-branes, alias extreme black holes. These
carry the scalar central charge in the type IIA supersymmetry algebra
(\ref{eq:onej}). Since these black hole solutions are extreme they saturate a
Bogomolnyi bound. Their mass is therefore a fixed multiple of their electric
charge and, because of the existence of the magnetic six-brane dual, this
electric charge is quantized. Hence their masses are quantized. Moreover,
because they saturate a Bogomolnyi bound the corresponding ground state
soliton supermultiplets are short ones of maximum spin 2. These are precisely
the features exhibited by the tower of Kaluza-Klein states obtained by
compactification of D=11 supergravity on $S^1$, and it is therefore natural to
conjecture that the D=10 type IIA extreme black hole states should be
{\it identified} as the KK states of D=11 supergravity (Townsend 1995a).
Alternatively, or perhaps equivalently, one can think of the KK states of
$S^1$ compactified D=11 supergravity as the {\it effective} description of the
black hole states of the D=10 IIA superstring theory (Witten 1995).
Another argument for the identification of the extreme black holes with KK
states is that the former can be interpreted as parallel plane waves
propagating at the speed of light in the compact direction (cf. Gibbons and
Perry 1984), so the corresponding quanta can be interpreted as massless
particles with momentum in the compact dimension, which is essentially a
description of KK modes.

Thus, the type IIA superstring is really an eleven-dimensional theory. From the
D=11 standpoint, the string coupling constant $g$ is $g=R^{2/3}$ (Witten 1995),
where $R$ is the radius of the 11th dimension, so that weak coupling
perturbation theory is a perturbation theory about $R=0$. This explains why the
critical dimension of the perturbative type IIA superstring theory is D=10. The
strong coupling limit is associated with the decompactification limit
$R\rightarrow\infty$ and D=11 supergravity can be interpreted as the effective
field theory at strong coupling (Witten 1995). However, the type IIA
superstring is really an 11-dimensional theory at {\it any} non-zero coupling,
weak or strong, and the question arises as to whether there is an intrinsically
11-dimensional description of this theory that is not merely an effective
one.

The only candidate for such a theory at present is the D=11 supermembrane but,
as noted earlier, its quantization via its worldvolume action leads to
difficulties. We can now explain why this should have been expected. While both
the extreme string solution of D=10 supergravity, in the string metric, and the
D=11 supermembrane solution of D=11 supergravity have a timelike singularity,
consistent with their interpretation as fundamental extended objects, the two
solutions differ in that the string singularity is naked whereas the membrane
singularity is `clothed'. The difference is significant. The fact that the
string singularity is naked shows that the string is classically structureless.
The worldsheet action is therefore an appropriate starting point for
quantization. In contrast, the supermembrane has a finite core due to its
horizon. Since the worldvolume action fails to take this classical structure
into account it is not an appropriate starting point for quantization. An
alternative approach (Townsend 1995a) that could circumvent this criticism
would be to quantize the classical membrane solution of D=11 supergravity. This
might run into difficulties caused by the singularity, however, in which case
some hybrid approach would be required. Clearly, there is much to do before we
can be sure whether a quantum 11-dimensional supermembrane makes physical
sense.

\vskip 1cm
\centerline{\bf References}
\bigskip

\noindent
Abraham E R C and Townsend P K, {\it Nucl. Phys.} {\bf B351} (1991)
313.

\noindent
de Azc{\' a}rraga J A, Gauntlett J P, Izquierdo J M and  Townsend P K,
{\it Phys. Rev. Lett.} {\bf 63} (1989) 2443.

\noindent
Bergshoeff E, Sezgin E and Townsend P K, {\it Phys. Lett.} {\bf 189B} (1987)
75.

\noindent
Callan C, Harvey J A and Strominger A, {\it Nucl. Phys.} {\bf B359} (1991);
{\it ibid} {\bf B367} (1991) 60.

\noindent
Campbell I C G and West P, {\it Nucl. Phys.} {\bf B243} (1984) 112.

\noindent
Cremmer E and Julia B, {\it Phys. Lett.} {\bf 80B} (1978) 48; {\it Nucl.
Phys.} {\bf B159} (1979) 141.

\noindent
Dabholkar A, Gibbons G W, Harvey J A and Ruiz-Ruiz F, {\it Nucl. Phys.} {\bf
B340} (1990) 33.

\noindent
de Wit B, L{\" u}scher M and Nicolai H, {\it Nucl. Phys.} {\bf B320} (1989)
135.

\noindent
Duff M J, Howe P S,  Inami T and Stelle K S, {\it Phys. Lett.} {\bf 191B}
(1987)
70.

\noindent
Duff M J and  Stelle K S, {\it Phys. Lett.} {\bf 253B} (1991) 113.

\noindent
Duff M J and Lu J X, {\it Nucl. Phys.} {\bf B354} (1991a) 141.

\noindent
Duff M J and Lu J X, {\it Phys. Lett.} {\bf 273B} (1991b) 409.

\noindent
Duff M J, Gibbons G W and Townsend P K, {\it Phys. Lett.} {\bf 332B} (1994)
321.

\noindent
Gauntlett J P, Harvey J A, and Liu J T, {\it Nucl. Phys. }{\bf B409}
(1993) 363.

\noindent
Gianni F and Pernici M, {\it Phys. Rev.} {\bf D30} (1984) 325.

\noindent
Gibbons G W and Perry M, {\it Nucl. Phys.} {\bf B248} (1984) 629.

\noindent
Gibbons G W, in {\sl  Supersymmetry, supergravity and related topics}, eds.
del Aguila F, de Azc\'arraga J A and Iba\~nez L E (World Scientific 1985) pp.
123-181.

\noindent
Gibbons G W, Kastor D, London L A J, Townsend P K and  Traschen J, {\it Nucl.
Phys.} {\bf B416} (1994) 850.

\noindent
Gibbons G W, Horowitz G T and Townsend P K, {\it Class. Quantum Grav.}
{\bf 12} (1995) 297.

\noindent
Gibbons G W and Townsend P K, {\sl Antigravitating BPS monopoles and dyons},
{\it Phys. Lett.} {\bf B}, {\it in press} (1995).

\noindent
Green M B and Schwarz J H, {\it Phys. Lett.} {\bf 136B} (1984) 367.

\noindent
G\"uven R, {\it Phys. Lett.} {\bf 276B} (1992) 49.

\noindent
Haag R, Lopusanski J and Sohnius M, {\it Nucl. Phys.} {\bf B88} (1975) 257.

\noindent
Horowitz G T and Strominger A, {\it Nucl. Phys.} {\bf B360} (1991) 197.

\noindent
Hull C M, {\sl String-string duality in ten dimensions}, hep-th/9506194
(1995).

\noindent
Hull C M and Townsend P K, {\it Nucl. Phys.} {\bf B438}, (1995a) 109.

\noindent
Hull C M and  Townsend P K, {\sl Enhanced gauge symmetries in
superstring theories}, {\it Nucl. Phys.} {\bf B}, {\it in press} (1995b).

\noindent
Harvey J A and Liu J T, {\it Phys. Lett.} {\bf 268B} (1991) 40.

\noindent
Khuri R R, {\it Nucl. Phys.} {\bf B387} (1992) 315.

\noindent
Mezincescu L, Nepomechie R I and Townsend P K, {\it Nucl. Phys.} {\bf B322}
(1989) 127.

\noindent
Nepomechie R I, {\it Phys. Rev.} {\bf D31} (1985) 1921.

\noindent
Osborn H, {\it Phys. Lett.} {\bf 83B} (1979) 321.

\noindent
Strominger A, {\it Nucl. Phys.} {\bf B343} (1990) 167.

\noindent
Teitelboim C, {\it Phys. Lett.} {\bf 167B} 1986, 63; {\it ibid} 69.

\noindent
Townsend P K, {\it Phys. Lett.} {\bf 202B} (1988) 53.

\noindent
Townsend P K, in {\sl Black Holes, Membranes, Wormholes and Superstrings}
eds. S. Kalara and D. Nanopoulos (World Scientific 1993) pp. 85-96.

\noindent
Townsend P K, {\it Phys. Lett.} {\bf 350B}, (1995a) 184.

\noindent
Townsend P K, {\it Phys. Lett.} {\bf 354B}, (1995b) 247.

\noindent
van Holten J W and Van Proeyen A, J. {\it Phys. A: Math. Gen.} {\bf 15} (1982)
3763.

\noindent
Weinberg S and Witten E, {\it Phys. Lett.} {\bf 96B} (1980) 59.

\noindent
Witten E and Olive D, {\it Phys. Lett.} {\bf 78B} (1978) 97.

\noindent
Witten E, {\it Nucl. Phys.} {\bf B443} (1995) 85.

\end{document}